\documentstyle[prl,aps]{revtex}

\begin{document}
\twocolumn[

\title{Entire Fredholm determinants for  Evaluation of\\
        Semi-classical and Thermodynamical Spectra}

\author{Predrag Cvitanovi\'c and G\'abor Vattay\cite{LAbs} }

\address{Niels Bohr Institute,
Blegdamsvej 17, DK-2100  Copenhagen \O, Denmark}

\date{\today}

\maketitle

\mediumtext
\begin{abstract}
Proofs that Fredholm determinants of
transfer operators for hyperbolic flows are entire
can be extended to a large new class of multiplicative
evolution operators. We construct such
operators both for the Gutzwiller semi-classical quantum mechanics
and for classical thermodynamic formalism,
and introduce a new functional determinant which is
expected to be entire for Axiom A flows, and whose zeros coincide with
the zeros of the Gutzwiller-Voros zeta function.
\end{abstract}
] 
\narrowtext

It has been established recently\cite{ER92,CR92,CRR93} that the
Gutzwiller-Voros zeta function\cite{voros} derived from the
Gutzwiller semi-classical trace formula\cite{gutbook}
is not an entire function.
In this letter we construct a new classical evolution operator
(\ref{tran_op}) whose
Fredholm determinant (\ref{main_res}) is entire for Axiom A flows,
and whose spectrum contains the semi-classical
Gutzwiller spectrum.  Many physically
realistic chaotic scattering systems\cite{gasp,eck} are of
Axiom A type, and for them the new determinant has much
better convergence properties than the Gutzwiller-Voros and Ruelle type
zeta functions utilized previously\cite{eck,GA}.

The main idea, extending the dynamical system to
the tangent space of the flow,
is suggested by one of the standard numerical methods for evaluation of
Lyapunov exponents\cite{BGS80}: start at $x_0$ with
an initial vector  ${\bf \xi}(0) $,
and let the flow transport it to ${\bf \xi}(t) $ along the
trajectory $x(t)= f^t(x_0)$.
The growth rate of this vector is multiplicative along the trajectory
\begin{equation}
\frac{| {\bf \xi}(t+t')|}{|{\bf \xi}(0)|}=
\frac{| {\bf \xi}(t+t')|}{|{\bf \xi}(t)|}\times
\frac{| {\bf \xi}(t)|}{|{\bf \xi}(0)|} \, ,
\label{mul_fl}
\end{equation}
and can be represented by
the trajectory of a ``unit" vector ${\bf u}(t) $ multiplied by
the factor $|{\bf \xi}(t)|/|{\bf \xi}(0)|$.
For asymptotic times and for almost every initial $(x_0,{\bf \xi}(0))$,
this factor converges to the leading eigenvalue of the
linearized stability matrix for the flow.

We implement this multiplicative evaluation
of stability eigenvalues by adjoining\cite{GRT}
the $d$-dimensional transverse tangent space
${\bf \xi}\in TU_x$, ${\bf \xi}(x) \cdot {\bf v}(x)=0$,
to the ($d$+1)-dimensional dynamical evolution
space $x\in U\subset {\bf R}^{d+1}$.
The dynamics in the $(x,\xi) \in U \times TU_x$ space
is governed by the system of
equations of variations\cite{arnold73}:
\[
\dot{x}={\bf v}(x) \,, \quad
\dot{{\bf \xi}}={\bf Dv}(x){\bf \xi }\, .
\]
Here ${\bf Dv}(x)$ is the transverse derivative matrix of the flow.
We write the solution as
\begin{equation}
x(t)=f^t(x_0) \,, \quad
{\bf \xi}(t)={\bf J}^t(x_0) \cdot {\bf \xi}_0 \, ,
\label{xit}
\end{equation}
with the tangent space vector ${\bf \xi}$ transported by
the transverse stability matrix\ ${\bf J}^t(x_0) = \partial x(t)/ \partial
x_0$.
In order to determine the length of the vector ${\bf \xi}$
we introduce a {\em signed norm},
a differentiable scalar function $g({\bf \xi})$
with the property $g(\Lambda {\bf \xi})=\Lambda g({\bf \xi})$
for any number $\Lambda $. While in general such norm is a
space dependent function $g({\bf \xi},x)$, we shall assume
here for reasons of notational simplicity that $g$ is a function of
${\bf \xi}$ only.
An example is the function
\begin{equation}
 g \left( \begin{array}{c}
        \xi_1 \\
        \xi_2 \\
        \cdots \\
        \xi_d
\end{array} \right)= \xi_d \,.
\label{proj_norm}
\end{equation}
Any vector ${\bf \xi}\in TU_x$ can now be represented by the product
$
{\bf \xi}=\Lambda {\bf u}
$,
where $ {\bf u}$ is a ``unit" vector in the sense that its
signed norm is $g({\bf u})=1$, and the factor
\begin{equation}
\Lambda^t(x_0,{\bf u}_0)=g(\xi(t)) = g({\bf J}^t(x_0) \cdot {\bf u}_0)
\label{lamb_def}
\end{equation}
is the multiplicative ``stretching'' factor introduced in (\ref{mul_fl})
\[
\Lambda^{t'+t}(x_0,{\bf u}_0)=\Lambda^{t'}(x(t),{\bf u}(t))
                                \, \Lambda^t(x_0,{\bf u}_0).
\]
The ${\bf u}$ evolution constrained to $ET_{g,x}$, the space of unit
tangent vectors transverse to the flow ${\bf v}$,
is given by rescaling of (\ref{xit}):
\begin{equation}
{\bf u}'
=R^t(x,{\bf u})=
\frac{1}{\Lambda^t(x,{\bf u})} {\bf J}^t(x) \cdot {\bf u} \, .
\label{R}
\end{equation}
Eqs. (\ref{xit}), (\ref{lamb_def}) and (\ref{R})
enable us to define a {\em multiplicative} evolution operator on
the extended space $ U \times ET_{g,x}$
\begin{equation}
{\cal L}^t({\bf u}',x';{\bf u},x)=
e^{h(x)} \delta(x'-f^t(x))
{\delta({\bf u}'-R^t(x,{\bf u})) \over |\Lambda^t(x,{\bf u})|^{\beta-1} }
\, ,
\label{tran_op}
\end{equation}
where
$h$ is a function additive along the trajectory, and $\beta$ is a number.
This should be contrasted to ``thermodynamic"\cite{ruelle,chicago5,budapest}
operators of form
\[
{\cal L}^t(x',x)=
e^{h(x)}
\delta(x'-f^t(x))
{1 \over |\Lambda^t(x)|^{\beta-1} }
\, ,
\]
with $\Lambda^t(x)$ an eigenvalue of ${\bf J}^t(x)$.
Such operators are {\em not} multiplicative in two or more transverse
dimensions,
for the simple reason that the eigenvalues of successive stability matrices
are in general not multiplicative
\[
\Lambda_{ab} \neq  \Lambda_a \Lambda_b.
\]
Here ${\bf J}_{ab}={\bf J}_{a}{\bf J}_{b}$ is the stability matrix\ of the
trajectory
consisting of consecutive segments $a$ and $b$, ${\bf J}_{a}$ and
${\bf J}_{b}$ are the stability matrices for these segments
separately, and $\Lambda$'s are their eigenvalues. In particular,
this lack of multiplicative property for $\Lambda$'s had until
now frustrated attempts\cite{CHAOS92} to construct evolution operators whose
spectrum contains the semi-classical Gutzwiller spectrum.

In order to derive the trace formula for the operator (\ref{tran_op})
we need to evaluate
$
\mbox{Tr}\, {\cal L}^t=\int dx d{\bf u}\, {\cal L}^t({\bf u},x;{\bf u},x)
$.
The $\int dx$ integral yields\cite{CEflows}
a weighted sum over primitive periodic orbits $p$
and their repetitions $r$:
\begin{eqnarray}
\mbox{Tr}\, {\cal L}^t &=&
\sum_p  T_p \sum_{r=1}^\infty
\frac{e^{rh_p}\delta(t-rT_p)}{\mid \det(1-{\bf J}_p^r)\mid}
\Delta_{p,r},
        \nonumber \\
\Delta_{p,r} &=& \int_g d{\bf u} \,
{ \delta({\bf u}-R^{T_p r}(x_p,{\bf u}))
  \over |\Lambda^{T_p r}(x_p,{\bf u})|^{\beta-1}       }
\, ,
\label{Tr_L}
\end{eqnarray}
where ${\bf J}_p$ is the prime cycle $p$ transverse stability matrix. As we
shall
see below, $\Delta_{p,r}$ is intrinsic to cycle $p$, and
independent of any particular cycle point $x_p$.

We note  next that if the trajectory $f^t(x)$ is periodic with period $T$,
the tangent space contains $d$ periodic solutions
\[
{\bf e}_i(x(T+t))={\bf e}_i(x(t)) \,, \quad i=1,...,d ,
\]
corresponding to the $d$ unit eigenvectors
$\{{\bf e}_1, {\bf e}_2, \cdots, {\bf e}_d\}$ of the transverse
stability matrix, with ``stretching" factors (\ref{lamb_def})  given by
its eigenvalues
\[
{\bf J}_p(x) \cdot {\bf e}_i(x) = \Lambda_{p,i} \, {\bf e}_i(x)
\,, \quad i=1,...,d.
\]
The $\int d{\bf u}$ integral in (\ref{Tr_L}) picks up contributions
from these periodic solutions.
In order to compute the stability of the $i$-th eigendirection
solution, it is convenient to expand the variation around
the eigenvector ${\bf e}_i$ in the stability matrix\ eigenbasis
$
\delta {\bf u} = \sum \delta u_\ell \, {\bf e}_\ell
\, .
$
The variation of the map (\ref{R}) at a complete period $t=T$ is then given by
\begin{eqnarray}
\delta R^T ({\bf e}_i)  &=&
\frac{{\bf J} \cdot \delta{\bf u}}{g({\bf J} \cdot {\bf e}_i)}
-\frac{{\bf J} \cdot {\bf e}_i }{g({\bf J} \cdot {\bf e}_i)^2}
\left(
 \frac{\partial g({\bf e}_i)}{\partial {\bf u}}
        \cdot {\bf J} \cdot \delta{\bf u}
\right)
                        \nonumber \\
        &=&
\sum_{k \neq i} {\Lambda_{p,k} \over \Lambda_{p,i }}
 \left( {\bf e}_k -{\bf e}_i\frac{\partial g({\bf
e}_i)}{\partial u_k}\right)\delta u_k
\, .
\label{var_R}
\end{eqnarray}
The $\delta u_i$ component does not contribute to this sum since
$g({\bf  e}_i+du_i{\bf  e}_i)=1+du_i$ implies
$\partial g({\bf e}_i) / \partial u_i = 1$.
Indeed, infinitesimal variations $\delta{\bf u}$ must satisfy
\[
g({\bf u}+\delta{\bf u})=g({\bf u})=1 \quad \Longrightarrow \quad
\sum_{\ell=1}^d \delta u_\ell
          { \partial g({\bf u}) \over \partial u_\ell} = 0
\,,
\]
so the allowed variations are of form
\[
\delta {\bf u} = \sum_{k\neq i}
                \left( {\bf e}_k -{\bf e}_i\frac{\partial g({\bf
e}_i)}{\partial u_k}\right)c_k \,, \quad |c_k| \ll 1
\, ,
\]
and in the neighborhood of the ${\bf e}_i$ eigenvector
the $\int {\bf u}$ integral can be expressed as
\[
\int_g d{\bf u} = \int \prod_{k\neq i} dc_k
\, .
\]
%
Inserting these variations into the $\int d{\bf u}$ integral we obtain
\begin{eqnarray}
\int_g d{\bf u}\, &&
\delta ( {\bf e}_i+\delta{\bf u} -R^{T}({\bf e}_i)
                                -\delta R^T({\bf e}_i) + \dots)
        \nonumber \\
= \, && \int \prod_{k\neq i} dc_k \,
        \delta \left((1-\Lambda_{k}/\Lambda_{i}) c_k
      + \dots\right)
        \nonumber \\
        &&= \prod_{k\neq i} { 1 \over \left| 1-
                        {\Lambda_{k} / \Lambda_{i } }\right|}
\, ,
\nonumber
\end{eqnarray}
and the $\int d{\bf u}$ trace (\ref{Tr_L}) becomes
\begin{equation}
\Delta_{p,r} =
\sum_{i=1}^d
\frac{1}{\mid \Lambda_{p,i}^r \mid^{\beta-1}}
\prod_{k\neq i}
\frac{1} {\mid 1-\Lambda^r_{p,k}/\Lambda^r_{p,i }\mid}
\, .
\label{del_pr}
\end{equation}
The corresponding Fredholm determinant
is obtained by observing\cite{CEflows} that
the Laplace transform of the trace
\[
\mbox{Tr} \, {\cal L}(s)=\int_{0_{+}}^{\infty}dt \, e^{st}\,\mbox{Tr} \,\,
{\cal L}(t)
\]
is a logarithmic derivative
$
\mbox{Tr} \, {\cal L}(s)=-\frac{d}{ds} \log F(s)
$
of the Fredholm determinant:
\begin{equation}
F(\beta,s)=\exp\left(-\sum_{p,r}\frac{e^{(h_p+sT_p)r}}
      {r \mid \det(1-{\bf J}_p^r)\mid}
        \Delta_{p,r}(\beta) \right).
\label{main_res}
\end{equation}
This determinant is the central result of this paper.
Its zeros correspond to the eigenvalues of the evolution
operator (\ref{tran_op}), and can be evaluated by standard cycle expansion
methods\cite{CRR93,AACI}.

In the ``thermodynamic'' formalism\cite{ruelle,chicago5,budapest}
for classical chaotic systems,
and in the Gutzwiller semi-classical description of systems with
chaotic classical counterpart \cite{gutbook},
$\beta$ is a parameter which plays the role of ``inverse temperature'',
and $h_p$ is the integral of some weight function $h(x)$
evaluated along the prime periodic orbit $p$.
In the semi-classical quantization case\cite{gutbook}
\begin{equation}
\beta=1/2, \quad
h_p=iS_p/\hbar +i \nu_p \pi/2, \quad s=0 \,,
\label{qm_act}
\end{equation}
where $S_p$ is the action of the periodic orbit,
and $\nu_p$ its Maslov index.
The classical correlation spectra are given by
$\beta=1$ and $h_p=0$.

The simplest application of (\ref{main_res}) is to
3-dimensional hyperbolic Hamiltonian flows (higher
dimensions need further evolution operators, for outer
products of vectors rather than single vectors). In this case
$\Lambda_1 = 1/\Lambda_2 = \Lambda$, and the Fredholm determinant
is given by
\begin{eqnarray}
F_{\sigma}(\beta,s)\,\,&=&\exp\left(-\sum_{p,r}
        \frac{\sigma_p^r}{r \mid \Lambda_p^r \mid}
      \frac{e^{r(h_p+sT_p)}}{(1-1/\Lambda_p^r)^2}
        \Delta_{p,r}(\beta)
\right)
                \nonumber \\
 \Delta_{p,r}(\beta) &=&
      \frac{\mid \Lambda_p^r \mid^{-\beta+1} }{1-1/\Lambda_p^{2r}}
      + \frac{ \mid \Lambda_p^r \mid^{\beta-3}}{1-1/\Lambda_p^{2r}}
\,.
\label{Ham_FD}
\end{eqnarray}
The extra multiplicative factor is set to
the eigenvalue sign $\sigma = \Lambda/|\Lambda|$ for $F_{-}$,
and to $\sigma =1$ for $F_{+}$; this will be used below.

The Gutzwiller-Voros zeta function corresponds to setting
$\Delta_{p,r} = |\Lambda^r_p|^{1/2} (1-1/\Lambda_p^r)$, and
the ``quantum Fredholm determinant''\cite{CR92} is obtained by
setting $\Delta_{p,r} = |\Lambda^r_p|^{1/2}$ in (\ref{Ham_FD}).
The practical advantage of (\ref{main_res}) over
the more familiar Gutzwiller-Voros and Ruelle type zeta functions
was demonstrated by detailed numerical studies\cite{CRR93}
of the related quantum Fredholm determinant\cite{CR92}.
In the systems studied, the quantum Fredholm determinant
appeared to be entire for all practical purposes; only the most
recent numerical investigations\cite{vattay}
reveal poles absent in (\ref{main_res}), see fig.~\ref{fig1}.

It can be shown\cite{vattay} that the Fredholm determinant
obtained by keeping only one of the terms in the sum in (\ref{del_pr})
is entire. This enables us to show that the Gutzwiller-Voros zeta
function $Z(E)$ for Axiom A flows is meromorphic in
the complex $E$ plane, as it can be written as a ratio of entire
functions; for 2-dimensional Hamiltonian systems
\begin{equation}
Z(E) = { F_{+}({1 \over 2},E) F_{-}({7 \over 2},E)
              \over
              F_{-}({3 \over 2},E) F_{+}({5 \over 2},E)  }
\,,
\label{Z_GV}
\end{equation}
where $F_{\sigma}$ includes only the first term in the
$\Delta_{p,r}$ sum (\ref{Ham_FD}).
The zeros of the Gutzwiller-Voros zeta function coincide
with the ones obtained from $F_{+}({1 \over 2},E)$.  Such relations
follow by inserting into $\Delta_{p,r}$ identities like
\[
1= {1 \over {1-1/\Lambda^r}}
   - {1 \over {\Lambda^r}} {1 \over {(1-1/\Lambda^r)}}
\, .
\]
($\sigma$ weight in (\ref{Ham_FD}) is needed to account for the
$1/\Lambda= \sigma/|\Lambda|$ term in the above indentity).

We illustrate a choice of the $g(\xi)$ function and
construction of the $R$ dynamics (\ref{R}) by a simple explicit example:
2-dimensional Hamiltonian dynamics reduced to
a 2-dimensional Poincar\'e section return map $x_{i+1}=f(x_{i})$.
The stability matrix\ of cycle $p$ is a product of the
[$2\times 2$] stability matrices
\[ {\bf J}_j= \left( \begin{array}{cc}
                      A_j & B_j \\
                      C_j & D_j
\end{array} \right)
\, ,
\]
where $A_j = \partial f_1(x_i) / \partial x_1$, and so on.
Assume the signed norm (\ref{proj_norm}) and multiply an
initial unit vector
by the first stability matrix\ in the product. The resulting vector can be
written as
\[ {\bf J}_1 \left( \begin{array}{c} \kappa_1 \\ 1 \end{array} \right)
= (C_1\kappa_1+D_1)
\left( \begin{array}{c}
        \frac{A_1\kappa_1+B_1}{C_1\kappa_1+D_1} \\
         1 \end{array} \right)
\, .
\]
Hence the dynamics acts on the unit vectors as a
rational fraction transformation
\begin{equation}
\kappa_{k+1}=R(x_k,\kappa_k)=
\frac{A_k\kappa_k+B_k}{C_k\kappa_k+D_k},
\label{racfrac}
\end{equation}
with the signed norm (\ref{lamb_def}) of the iterated vector given by
\[\Lambda^{n_p}(x_{1},\kappa_{1}) =
\prod_{i=1}^{n_p}(C_i\kappa_i+D_i)
\, .
\]
(In the case of 2-dimensional billiards,
$\kappa$  is the Bunimovich-Sinai curvature\cite{BS80}).
For a periodic orbit $\kappa_{n_p}=\kappa_0$,
the unit vector is an eigenvector of
the stability matrix, and the corresponding eigenvalue is
$\Lambda_p=\prod 
 (C_i\kappa_i+D_i)$.
For 3-dimensional hyperbolic flows
there are two distinct $\kappa_{p,i}$ values,
one for the expanding, and the other for the contracting eigenvalue.
The derivative of $R^{n_p}$ is easily evaluated;
\[
{\partial \over \partial \kappa} {R^{n_p}}(\kappa_1,x_1)
=\prod_{i=1}^{n_p}R'(\kappa_i,x_i)
=\prod_{i=1}^{n_p}\frac{ 1 }{(C_i\kappa_i+ D_i)^2}
\,,
\]
(we have used $\det({\bf J}_i)=1$)
and we again obtain the trace formula (\ref{Ham_FD}).

In conclusion, we have constructed a classical evolution
operator for semi-classical quantization, and derived a
new determinant for periodic orbit quantization of chaotic
dynamical systems.
The main virtue of the determinant (\ref{main_res})
is that the theorem of H.H.~Rugh\cite{Rugh92}, applicable to multiplicative
evolution operators such as (\ref{tran_op}), implies that this
determinant should be entire for Axiom A flows,
{\em ie.} free of poles in the entire complex $s$ or complex energy plane.
One consequence of this general result is that the cycle
expansions of the new Fredholm determinant
should converge faster than exponentially.
Our numerical tests support above claims\cite{vattay}; one
example is given in fig.~\ref{fig1} which demonstrates that the
new determinant is analytic to the limit of numerical precision
of current cycle expansions, and well beyond both
the Gutzwiller-Voros zeta function and the quantum Fredholm determinant.

\acknowledgements

G.V. is indebted to D. Sz\'asz and A. Kr\'amli for pointing
out the importance of the Bunimovich-Sinai curvatures, to
the Sz\'echenyi Foundation and OTKA F4286 for the support, and
to the Center for Chaos and Turbulence Studies, Niels Bohr Institute, for
hospitality. P.C. thanks the Carlsberg Fundation for support.

\begin{figure}
\caption[]{
The leading semi-classical resonances in the $k$ complex wave-number plane
($\times$) for the determinant (\ref{main_res}), compared with
($\Box$) the zeros of the ``quantum Fredholm determinant"\cite{CR92,CRR93}.
While the quantum Fredholm determinant has a finite region of analyticity
(the bottom line of zeros
reflects a pole expected at $Im(k) = -4.559843\dots$, indicated
by the dashed line labeled ``QF"),
the new determinant shows no numerical indication of any poles,
and enables us to reach resonances deeper down in the complex plane.
The Gutzwiller-Voros determinant is reliable only down to the line
labeled ``GV",
Im$(k) = -2.491905\dots$, the upper bound on the poles of
determinant  $F_{-}({3 \over 2},E)$ in (\ref{Z_GV}).
The dynamical system tested is the Hamiltonian H\'enon map
$x_{k+1}=1-ax^2_k-x_{k-1}$
at $a=20$, with cycle expansions truncated to cycles up to period $18$.
We take as action $S_p/\hbar =k n_p$, and as the Maslov phase $\nu_p ={2 n_p}$
(this is a normal-form model 
of a 3-disc repeller, see ref.~\cite{CRR93}).
}
\label{fig1}
\end{figure}
\end{document}